\documentclass{iau}
\usepackage{graphicx} 

\title[IAUS291.~~Formation of the planet orbiting PSR J1719--1438] 
{Formation of the planet orbiting the millisecond pulsar J1719--1438} 

\author[L.~M.~van~Haaften et al.]  
{L.~M.~van~Haaften$^{1}$,
 G.~Nelemans$^{1,2}$,
 R.~Voss$^1$
 \and P.~G.~Jonker$^{3,1,4}$}

\affiliation{$^1$Department of Astrophysics/ IMAPP, Radboud University Nijmegen, P.O. Box 9010, 6500 GL Nijmegen, The Netherlands, \\ email: {\tt L.vanHaaften@astro.ru.nl} \\[\affilskip]
$^2$Institute for Astronomy, KU Leuven, Celestijnenlaan 200D, 3001 Leuven, Belgium \\[\affilskip]
$^3$SRON, Netherlands Institute for Space Research, Sorbonnelaan 2, 3584 CA, Utrecht, The Netherlands \\[\affilskip]
$^4$Harvard-Smithsonian Center for Astrophysics, 60 Garden Street, Cambridge, MA 02138, USA}

\pubyear{2012}
\volume{291}  
\jname{\mbox{Neutron Stars and Pulsars: Challenges and Opportunities after 80 years}}
\editors{J. van Leeuwen, ed.} 
\begin{document}

\maketitle

\begin{abstract}
In 2011, Bailes \etal\ reported on the discovery of a detached companion in a 131 minute orbit around PSR J1719--1438, a 173 Hz millisecond pulsar. The combination of the very low mass function and such a short orbital period is unique. The discoverers suggested that the progenitor system could be an ultracompact X-ray binary (UCXB), which is a binary with a sub-hour orbital period in which a (semi-)degenerate donor fills its Roche lobe and transfers mass to a neutron star. The standard gravitational-wave driven UCXB scenario, however, cannot produce a system like PSR J1719--1438 as it would take longer than the age of the Universe to reach an orbital period of 131 min. We investigate two modifications to the standard UCXB evolution that may resolve this discrepancy. The first involves significant heating and bloating of the donor by pulsar irradiation, and in the second modification the system loses orbital angular momentum via a fast stellar wind from the irradiated donor, additional to the losses via the usual gravitational wave radiation. In particular a donor wind is effective in accelerating orbital expansion, and even a mild wind could produce the 131 minute period within the age of the Universe.
We note that UCXBs could be an important class of progenitors of solitary millisecond radio pulsars.

\keywords{pulsars: individual (PSR J1719--1438), stars: mass loss, binaries: close, planets and satellites: formation, X-rays: binaries}
\end{abstract}


\firstsection 
\section{Introduction}

PSR J1719--1438, with a 5.8 ms spin period, was found to be orbited every 131 min by a companion \cite[(Bailes \etal\ 2011)]{bailes2011}. The mass function of $\sim 10^{-9}\ M_{\odot}$ implies that in the (a priori) very likely case that we do not observe the PSR J1719--1438 system close to face-on, the companion must have a mass of the order of $\sim \! 10^{-3}\ M_{\odot}$.
This system is very interesting because almost all millisecond radio pulsars either have a stellar-mass companion or no companion at all. There are black widow X-ray binaries with similar orbital periods but those are semi-detached and have much more massive companions.
The system is detached, and from its orbital period a minimum average density of the companion of $23\ \mbox{g cm}^{-3}$ can be derived.

\section{Composition of the companion}

 \begin{figure}[b]
 \begin{center}
  \vspace{-5mm}\includegraphics[width=3.8in]{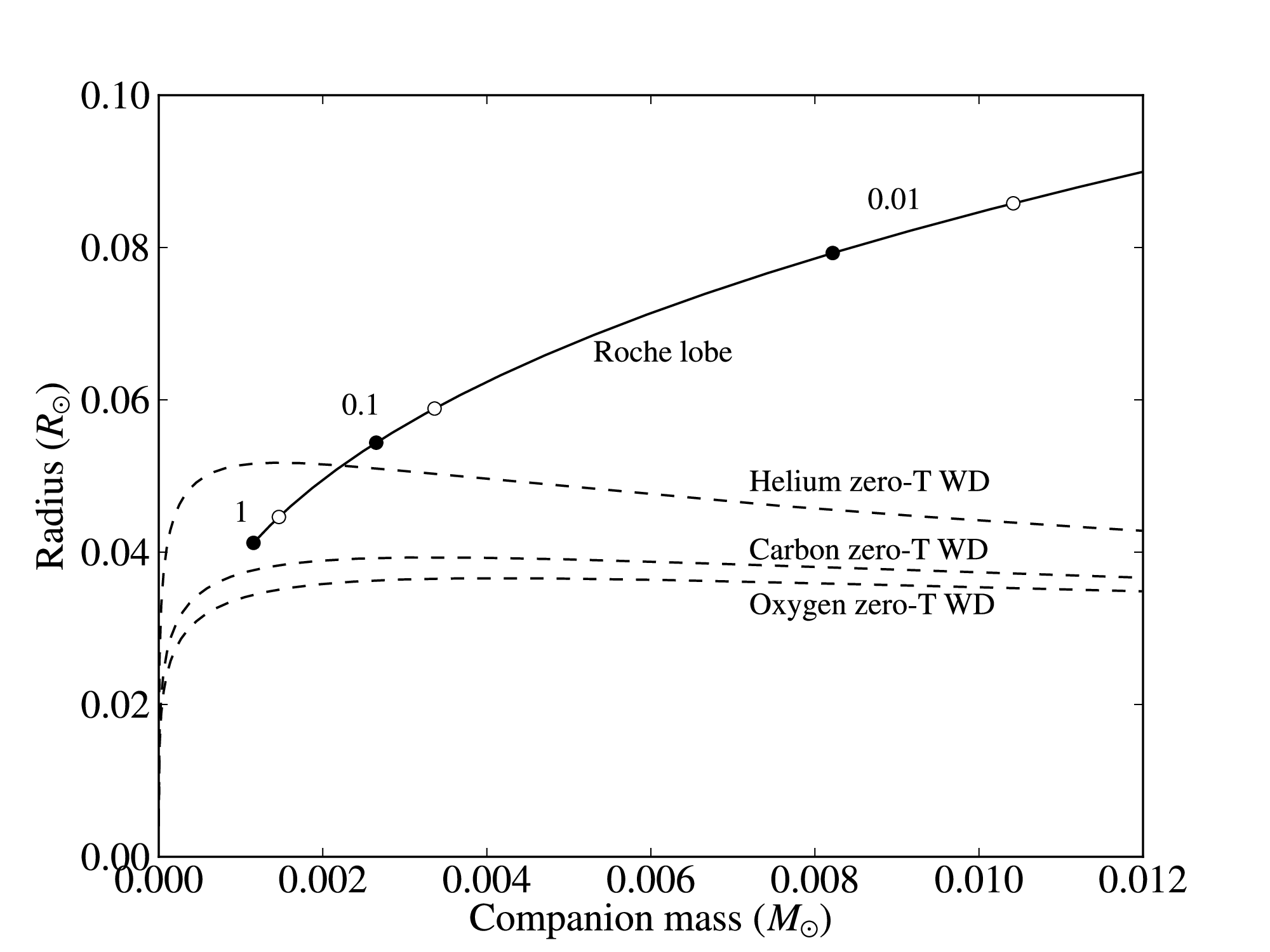} 
  \caption{Mass-radius relations for zero-temperature white dwarfs \cite[(Rappaport \etal\ 1987)]{rappaport1987} for pure helium, carbon and oxygen (dashed), and the relation between the Roche-lobe radius and the mass of the companion of PSR J1719--1438 (solid). Filled circles ($1.4\ M_{\odot}$ neutron star) and open circles ($2\ M_{\odot}$ neutron star) indicate the 1, 0.1 and 0.01 a priori probabilities that the companion mass is higher than indicated. Figure from \cite{vanhaaften2012j1719}.} 
    \label{fig:wdrad}
 \end{center}
 \end{figure}

In Fig.~\ref{fig:wdrad} we show the constraints on the companion set by its minimum density (i.e., its Roche-lobe radius at a given mass, which is the maximum radius), and the radii of zero-temperature white dwarfs of different compositions, which give the minimum radius. We see that a carbon-oxygen composition is the most likely, followed by helium, but that requires an inclination with an a priori probability near $14\%$, or less in case of a finite-temperature companion. Hydrogen is possible if we see the systems nearly face-on, with an a priori probability near $0.1\%$

\section{Ultracompact X-ray binary scenario}

\cite{bailes2011} suggested that this system could have evolved from an ultracompact X-ray binary (UCXB), which consists of a (semi-)degenerate donor transferring mass to a neutron star, driven by angular momentum loss via gravitational wave radiation \cite[(e.g. Savonije \etal\ 1986)]{savonije1986}. The mean donor density is one-to-one related to the orbital period, as it always the case for a Roche-lobe filling star with a more massive companion. Therefore, as the donor loses mass and becomes less dense, the orbit expands, and the mass transfer rate decreases. However, UCXBs are expected to reach orbital periods of at most 90 min within the age of the Universe \cite[(Deloye \& Bildsten 2003)]{deloye2003}.

We discuss two modifications to this standard UCXB scenario that could explain how they could reach an orbital period of 131 min. The first is a relatively large donor size, the second is a stellar wind from the donor. Both are caused by heating and irradiation of the donor by the accretion disk, accretor, and magnetosphere.

 \begin{figure}[b]
 \begin{center}
  \vspace{-5mm}\includegraphics[width=3.8in]{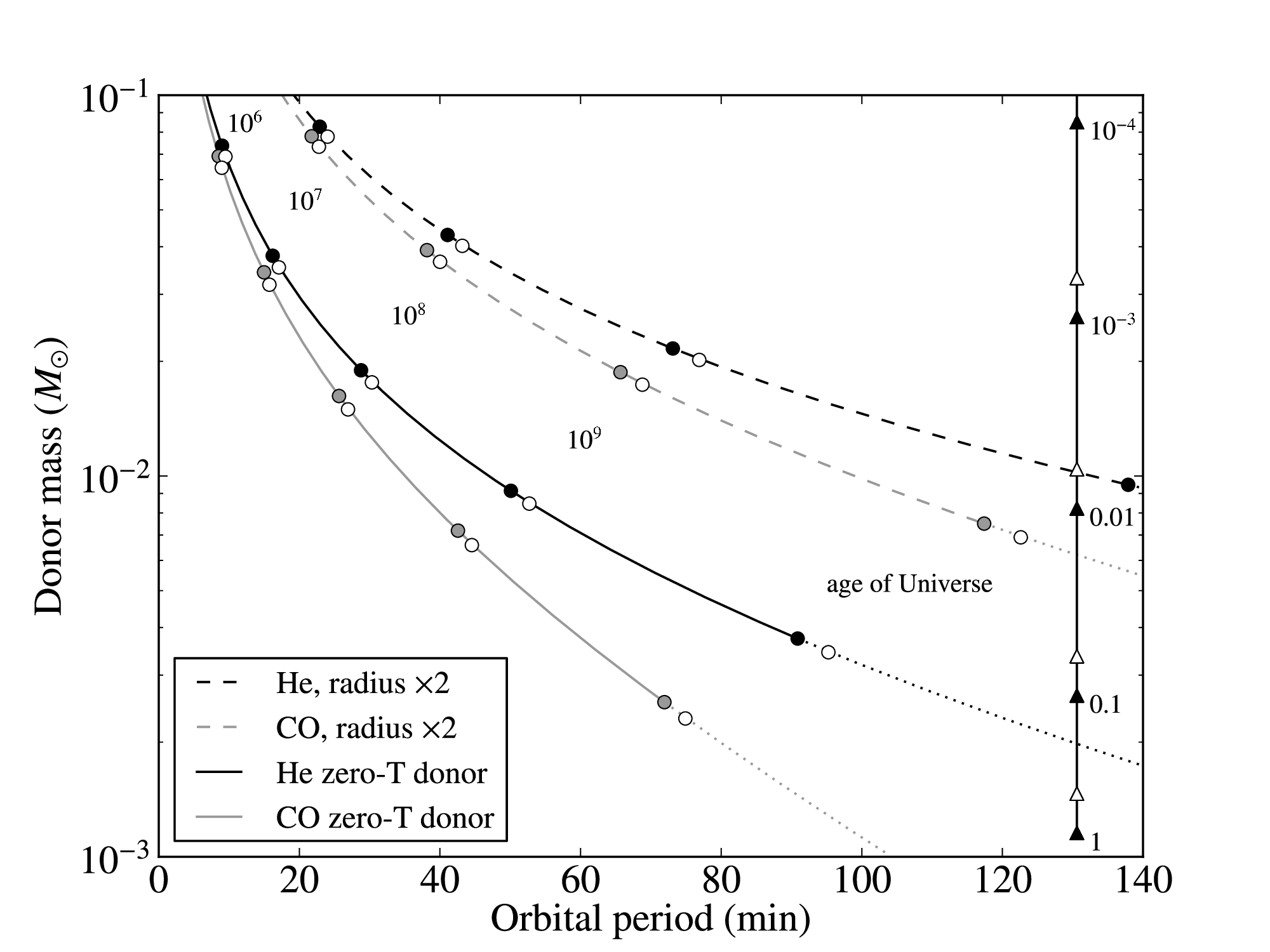} 
  \caption{Donor mass versus orbital period for UCXBs based on tracks by \cite{vanhaaften2012evo}. The solid curves show the evolution of UCXBs with zero-temperature white dwarf donors. The dashed curves show the same except that the donors have twice as large radii at all masses. The circles and numbers mark ages in yr. Filled (open) symbols correspond to an initial accretor mass of $1.4\ M_{\odot}\ (2\ M_{\odot})$. The dotted parts of the curves take longer than the age of the Universe to reach. The vertical line is located at the orbital period of PSR J1719--1438, and the triangles mark the a priori probabilities that the donor mass is higher than indicated, based on the mass function. Figure adapted from \cite{vanhaaften2012j1719}.} 
    \label{fig:mp}
 \end{center}
 \end{figure}

\section{Increased donor size}
The size of the donor in an UCXB influences the mass transfer rate. A larger (`bloated') donor with respect to the zero-temperature radius implies a higher mass is required to have a given average density, which corresponds to an orbital period, as mentioned before. A higher donor mass in turn implies a higher mass transfer rate and faster evolution. In Fig.~\ref{fig:mp} is shown that UCXBs with relatively large donors can reach a period of 131 within the age of the Universe. However, the donor mass would be about ten times larger than suggested by the mass function. This is unlikely to be the main explanation of the 131 min period, but it could contribute.

\section{Stellar wind from the donor}

 \begin{figure}[t]
 \begin{center}
  \includegraphics[trim=0 0 0 30, clip, width=3.8in]{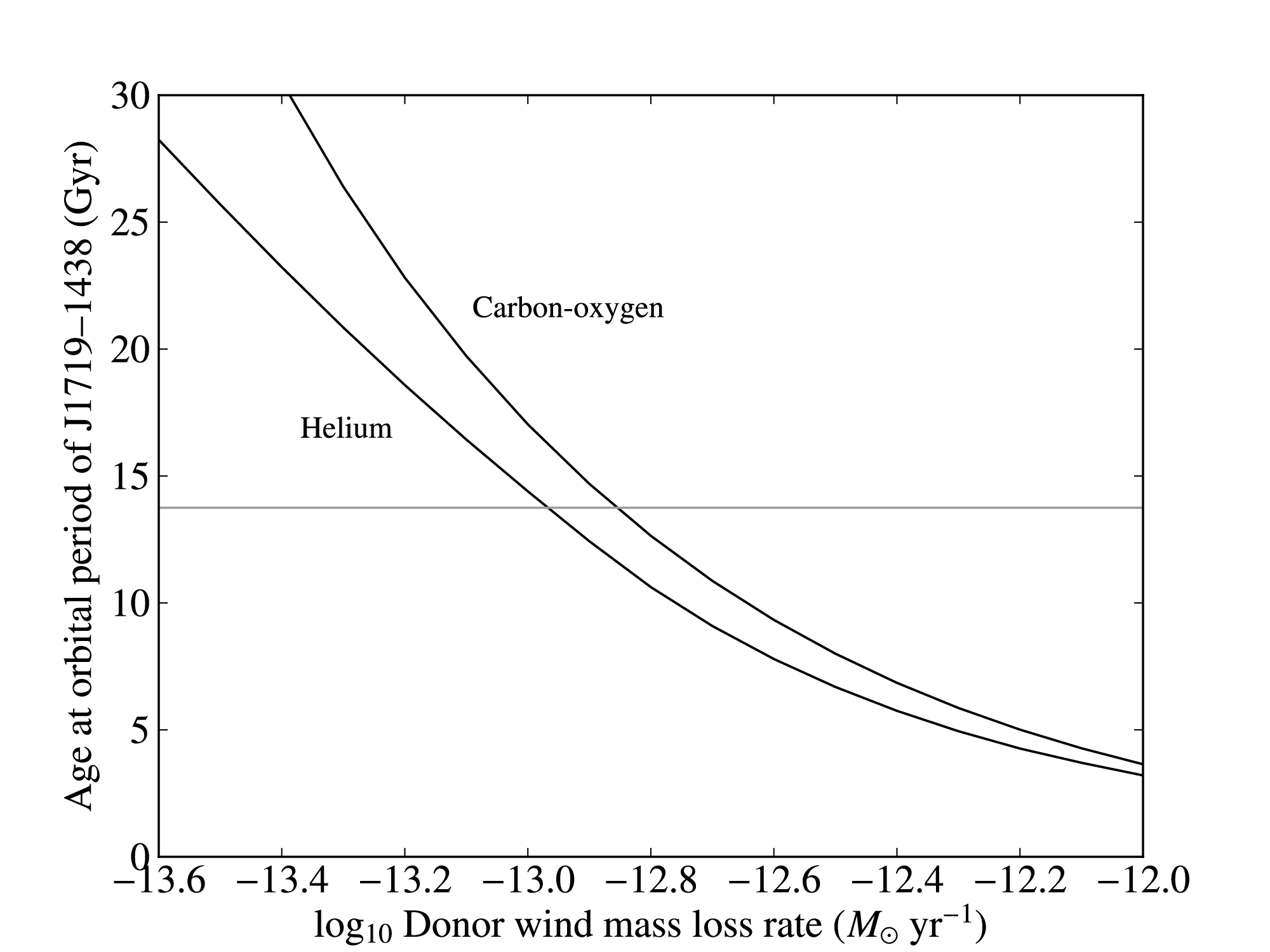} 
  \caption{Age of an UCXB when its orbital period reaches the orbital period 131 min of PSR J1719--1438 versus donor wind rate. The horizontal line shows the age of the Universe. Figure from \cite{vanhaaften2012j1719}.} 
    \label{fig:time}
 \end{center}
 \end{figure}

In a binary, the specific angular momentum is highest in the lowest-mass component. At the extreme mass ratio found in UCXBs, a large amount of angular momentum can be lost by systems if the donor loses matter in a fast, isotropic wind. Such a wind can be caused by irradiation of the donor. Fig.~\ref{fig:time} shows that a relatively low wind mass loss rate during the entire evolution is already enough to speed the orbital expansion to a period of 131 min within the age of the Universe.

Long-term observations by the \textit{Rossi XTE} All-Sky Monitor show that most of the UCXBs with orbital periods longer than 40 min are one to two orders of magnitude brighter than expected when their mass transfer was driven only by angular momentum loss via gravitational wave radiation \cite[(van Haaften \etal\ 2012c)]{vanhaaften2012asm}. Also, \cite{knigge2011} found that Cataclysmic Variables below the period gap evolve faster than expected from gravitational wave emission only. This suggests that they indeed evolve faster, which supports the hypothesis that PSR J1719--1438 is an UCXB descendant.

Strong evidence comes from the PSR J1311--3430 system \cite[(Pletsch \etal\ 2012)]{pletsch2012}, an UCXB with an unusually long orbital period of 93.8 min \cite[(Romani 2012, Kataoka \etal\ 2012)]{romani2012,kataoka2012} and an evaporating helium donor \cite[(Romani \etal\ 2012)]{romanietal2012}.
\cite{benvenuto2012} explained the formation of PSR J1719--1438 via an originally hydrogen-rich donor star.

The present detached state of the system may be caused by the gradually decreasing irradiation, which could result in the donor shrinking inside its Roche lobe. Then the accretion rate would decrease, further decreasing the level of donor irradiation.

\section{Conclusion}

PSR J1719--1438 very likely has a low-mass companion, composed mainly of carbon/oxygen or possibly helium. Its orbital period of 131 min could have been reached by an ultracompact X-ray binary whose late-time evolution was driven by angular momentum loss via a stellar wind from the donor, rather than the emission of gravitational waves, which subsequently became detached. A wind of $\gtrsim 3 \times 10^{-13}\ M_{\odot}\ \mbox{yr}^{-1}$ would be sufficient. The composition of the companion is most likely carbon-oxygen, though helium cannot be ruled out.

Old UCXBs are expected to have very low donor masses (below $\sim \! 0.01\ M_{\odot}$). If PSR J1719--1438 has indeed formed out of an UCXB, then we would expect to see more binary millisecond radio pulsars with very low mass companions. Such systems are very rare though, whereas solitary millisecond radio pulsars are relatively common. This suggests the possibility that low-mass companions of UCXBs are completely evaporated \cite[(Ruderman \etal\ 1989)]{ruderman1989}, leaving a population of solitary millisecond radio pulsars.


\begin{thebibliography}{}

\bibitem[Bailes \etal\ (2011)]{bailes2011}
{Bailes, M., Bates, S.~D., Bhalerao, V., et al.} 2011,
\textit{Science}, 33, 1717

\bibitem[Benvenuto \etal\ (2012)]{benvenuto2012}
{Benvenuto, O. G., De Vito, M. A., \& Horvath, J. E.} 2012,
\textit{ApJ}, 753, L33

\bibitem[Deloye \& Bildsten (2003)]{deloye2003}
{Deloye, C. J., \& Bildsten, L.} 2003,
\textit{ApJ}, 598, 1217

\bibitem[Kataoka \etal\ (2012)]{kataoka2012}
{Kataoka, J., Yatsu, Y., Kawai, N., et al.} 2012,
\textit{ApJ}, 757, 176

\bibitem[Knigge \etal\ (2011)]{knigge2011}
{Knigge, C., Baraffe, I., \& Patterson, J.} 2011,
\textit{ApJS}, 194, 28

\bibitem[Pletsch \etal\ (2012)]{pletsch2012}
{Pletsch, H. J., Guillemot, L., Fehrmann, H., et al.} 2012,
\textit{Science}, doi:10.1126/science.1229054

\bibitem[Rappaport \etal\ (1987)]{rappaport1987}
{Rappaport, S., Ma, C. P., Joss, P. C., \& Nelson, L. A.} 1987,
\textit{ApJ}, 322, 842

\bibitem[Romani (2012)]{romani2012}
{Romani, R. W.} 2012,
\textit{ApJ}, 754, L25

\bibitem[Romani \etal\ (2012)]{romanietal2012}
{Romani, R. W., Filippenko, A. V., Silverman, J. M., et al.} 2012,
\textit{ArXiv e-prints}, 1210.6884v1

\bibitem[Ruderman \etal\ (1989)]{ruderman1989}
{Ruderman, M., Shaham, J., \& Tavani, M.} 1989,
\textit{ApJ}, 336, 507

\bibitem[Savonije \etal\ (1986)]{savonije1986}
{Savonije, G. J., de Kool, M., \& van den Heuvel, E. P. J.} 1986,
\textit{A\&A}, 155, 51

\bibitem[van Haaften \etal\ (2012a)]{vanhaaften2012j1719}
{van Haaften, L.~M., Nelemans, G., Voss, R., \& Jonker, P.~G.} 2012a,
\textit{A\&A}, 541, A22

\bibitem[van Haaften \etal\ (2012b)]{vanhaaften2012evo}
{van Haaften, L.~M., Nelemans, G., Voss, R., et al.} 2012b,
\textit{A\&A}, 537, A104

\bibitem[van Haaften \etal\ (2012c)]{vanhaaften2012asm}
{van Haaften, L.~M., Voss, R., \& Nelemans, G.} 2012c,
\textit{A\&A}, 543, A121





\end{thebibliography}
\end{document}